\documentclass[twocolumn,showpacs,preprintnumbers,amsmath,amssymb,prl]{revtex4}

\usepackage{dcolumn}
\usepackage{bm}
\usepackage{hyperref} 
\usepackage{amsmath}
\usepackage{amsfonts}     
\usepackage{graphics}  
\usepackage{psfrag} 
\usepackage{graphicx} 
\usepackage{epsfig}    
\usepackage{rotating}  
 \usepackage[latin1]{inputenc}
 \usepackage{color}
 \usepackage{rotating}

\begin{document}

\preprint{LMU-ASC 12/08}

\title{Instability of spatial patterns and its ambiguous impact on species diversity}

\author{Tobias Reichenbach  and Erwin Frey}
\affiliation{Arnold Sommerfeld Center for Theoretical Physics (ASC) and
  Center for NanoScience (CeNS), Department of Physics,
  Ludwig-Maximilians-Universit\"at M\"unchen, Theresienstrasse 37,
  D-80333 M\"unchen, Germany}


\date{\today}
\begin{abstract}
Self-arrangement of individuals into spatial patterns often accompanies and promotes species diversity in ecological systems.  Here, we investigate pattern formation arising from cyclic dominance of three species, operating near a bifurcation point. In its vicinity, an Eckhaus instability occurs, leading to convectively unstable ``blurred'' patterns.  At the bifurcation point, stochastic effects dominate  and induce counterintuitive effects on diversity:  Large patterns, emerging for medium values of individuals' mobility, lead to rapid species extinction, while small 
 patterns (low mobility) promote diversity, and high mobilities render spatial structures irrelevant. We provide a quantitative analysis of these phenomena, employing a complex Ginzburg-Landau equation.
\end{abstract}

\maketitle

Biodiverse ecosystems comprise complex interactions of a large number of individuals and species leading to rich spatio-temporal community structures~\cite{May}.
Much efforts in theoretical and biological physics are currently devoted to qualitative and quantitative understanding of basic mechanisms that maintain their diversity. Hereby,
within exemplary models, the formation of dynamic spatial patterns has been identified as a key promoter~\cite{levin-1974-108,hassell-1994-370,durrett-1994-46,reichenbach-2007-448}.
In particular, crucial influence of self-organized patterns on biodiversity has been demonstrated in recent experimental studies~\cite{kerr-2002-418}, employing three bacterial strains that display cyclic competition. The latter is metaphorically described by the game `rock-paper-scissors' where rock smashes scissors, scissors cut paper, and paper wraps rock in turn.
Such non-hierarchichal  dynamics  has also been found in, e.g.,  lizard populations in California~\cite{sinervo-1996-340} and coral reef invertebrates~\cite{jackson-1975-72}.
For the three bacterial strains, and 
for low microbes' motility, cyclic dominance leads to stable coexistence of all three strains through self-formation of spatial patterns~\cite{kerr-2002-418}. In contrast, stirring the system, as  can also result from high motilities of the individuals, destroys the spatial structures and  the strain coexistence~\cite{kerr-2002-418}.

Here, from theoretical studies, we show that cyclic competition of species  can lead to highly non-trivial spatial patterns as well as counterintuitive effects  on biodiversity. Investigating cyclic dynamics near a (degenerate) bifurcation, we find that pattern formation is only weak and mainly influenced by stochastic effects. Namely, in a prototypical model where three species compete cyclically, and the deterministic rate equations (RE) for the densities' time evolution predict weakly instable coexistence (near or at neutral stability), we demonstrate the generic formation of convectively instable spiral waves. The instability of the latter appears as a ``blurring'' of the spirals and is the stronger the closer the RE operate near the bifurcation point, i.e.~near the parameter point where the RE predict neutral stability. Consequently, patterns take shape only weakly and allow for major influence of stochasticity. This effect is most pronounced at the bifurcation point; there, noise remains as the only source of pattern formation.
Furthermore, at the bifurcation point, we uncover a counterintuitive \emph{destabilizing} effect of patterns on the stability of coexistence.  Similar to what has been found in other studies, see Refs.~\cite{szabo-2002-65,kerr-2002-418,mobilia-2006-73,szabo-2007-446,reichenbach-2007-448} and references therein,
for low individuals' mobility, the size of the emerging spatial structures lies  much below the size of the ecosystem and allows  for stable coexistence of all species. However, we show that a mobility exceeding a first critical value leads to rapid extinction of species with only one surviving. This scenario is connected to weakly formed patterns that span over the whole system, revealing the antagonistic character of such large patterns on  biodiversity. Only when mobility lies above a second threshold, a third, effectively well-mixed regime emerges. There, patterns and spatial correlations do not form and the fate of species diversity
is determined by the character of the species competitions alone.

We design a spatial and stochastic  model of cyclically competing populations in the following way. Individuals of three species $A$, $B$, and $C$ populate  a square lattice, such that each site is occupied by one individual or left empty. Each agent can interact with its four nearest neighbors by either exchanging positions at a rate $\epsilon$ (individuals are mobile), or by cyclic competition. The latter interactions are described in the language of chemical reactions; generically, we consider:\\
\hspace*{-0.7cm}
\parbox{3.2cm}{
\begin{eqnarray}
&AB\stackrel{1}{\longrightarrow} AA,   \cr
&BC\stackrel{1}{\longrightarrow} BB,   \cr
&CA\stackrel{1}{\longrightarrow} CC,    \label{react1}  
\end{eqnarray}
}\hfill
\hspace*{-0.4cm}
\parbox{3.2cm}{
\begin{eqnarray}
&AB\stackrel{\gamma}{\longrightarrow}  A\oslash,   \cr
&BC\stackrel{\gamma}{\longrightarrow}  B\oslash,  \cr
&CA\stackrel{\gamma}{\longrightarrow}  C\oslash,  \label{react2}  
\end{eqnarray}
}\hfill
\hspace*{-0.4cm}
\parbox{3.2cm}{
\begin{eqnarray}
&A\oslash\stackrel{\mu}{\longrightarrow} AA,   \cr
&B\oslash\stackrel{\mu}{\longrightarrow} BB,  \cr
&C\oslash\stackrel{\mu}{\longrightarrow} CC.  \label{react3}  
\end{eqnarray}
}\\
Reactions~(\ref{react1}) describe consumption of an individual by another of a predominant species, and immediate reproduction of the latter. Cyclic dominance appears as $A$ consumes $B$ and reproduces, while $B$ preys on $C$ and $C$ feeds on $A$ in turn. We consider the most symmetric version where all species are symmetric, and fix the time-scale by setting the rates for these reactions to one.  Reactions~(\ref{react2}) encode solely consumption, leaving an empty site $\oslash$. These reactions occur at a rate $\gamma$, and are decoupled from reproduction, Eqs.~(\ref{react3}), which happens at a rate $\mu$. Note that reactions~(\ref{react1}) and (\ref{react3}) describe two different mechanisms of reproduction, both of which are important for  ecological systems:  In (\ref{react1}),  an individual reproduces when having consumed a prey, due to thereby  increased fitness.   In contrast, in reactions~(\ref{react3}) reproduction depends solely on the availability of empty space.

\begin{figure}  
\begin{center}
\includegraphics[scale=1]{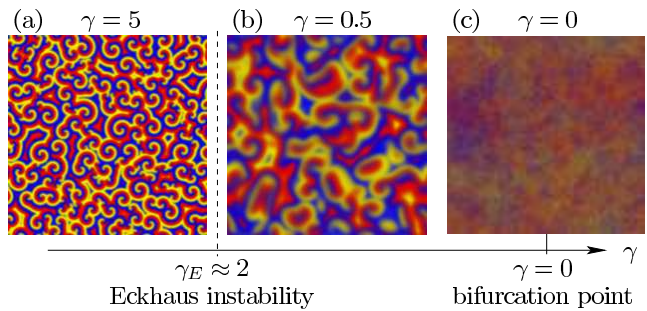}
\caption{
Snapshots of the biodiverse state for $D=1\times 10^{-5}$. (a), For large rates $\gamma$,  entangled and stable spiral waves form. (b), A convective (Eckhaus) instability occurs at  $\gamma_E\approx 2$; below this value, the spiral patterns blur.  ({c}), At the bifurcation point $\gamma=0$, only very weak  spatial modulations  emerge;  we have amplified them by a factor two for better visibility. The snapshots stem from numerical solution of the SPDE~(\ref{SPDE}) with initially homogeneous  densities $a=b=c=1/4$. \label{snapshots}}
\end{center}           
\end{figure} 
The RE encode the deterministic behavior of these reactions in the well-mixed scenario. Dependent on the type of reactions, they can  yield stable, unstable, or neutrally stable diversity. For the above defined model, with $\vec{x}=(a,b,c)$ denoting the densities of the three distinct species and $\rho=a+b+c$ the total density, the RE read,
\begin{equation}
\partial_t x_i = x_i\left[\mu(1-\rho)  + x_{i+1} -(1+\gamma) x_{i+2} \right]\equiv \mathcal{A}_i\,.
\label{RE}
\end{equation}
Hereby, the indices are understood as modulo $3$.
These equations have been analyzed by May and Leonard~\cite{may-1975-29} 
and also appear in theoretical descriptions of the Kuppers-Lortz instability~\cite{busse-1980-208,yuhai-1992-69}. Generically, Eqs.~(\ref{RE}) possess an unstable internal fixed point as well as heteroclinic orbits, where the system cycles between states with nearly only one species, leading  to rapid extinction of all but one species when stochastic effects are included. A degenerate bifurcation emerges at $\mu=\gamma=0$, i.e.~when only the reactions~(\ref{react1}) are present. In this situation, neutrally stable orbits surround an internal fixed point, being neutrally stable as well~\cite{may-1975-29,reichenbach-2006-74}. Finite-size fluctuations invalidate the neutral stability and induce extinction of two species after a characteristic time $T$ which is proportional to the system size $N$~\cite{reichenbach-2006-74}. In the following, we investigate the system's behavior in the vicinity of this bifurcation point; for illustration, we consider equal selection and reproduction rates  $\gamma=\mu$.

In the spatial model, mobility of individuals, stemming from random local exchanges, leads to their diffusion on the lattice
with a diffusion constant  $D=2\epsilon N^{-1}$~\cite{Redner,reichenbach-2007-99}. Employing a continuum limit of large systems, $N\to\infty$ with diffusivity $D$ kept fixed, the stochastic spatial system becomes describable by stochastic partial differential equations (SPDE), see References~\cite{reichenbach-2007-99}; they read
\begin{eqnarray}
\partial_t x_i&=& D\nabla^2x_i + \mathcal{A}_i + \sum_j \mathcal{C}_{ij}\xi_j\,. \label{SPDE}
\end{eqnarray}  
In these equations, a term $D\nabla^2$ describes individuals' diffusion on the lattice; it reveals that the size of possible spatial structures is proportional to $\sqrt{D}$~\cite{reichenbach-2007-99}. The nonlinear terms $\mathcal{A}_i$ coincide with those of the RE~({\ref{RE}}) and take interactions into account.
Noise terms, stemming from a system-size expansion~\cite{Gardiner,traulsen-2005-95} are added which scale as the square root of $N$. In Eqs.~(\ref{SPDE}), $\xi_i$ denote uncorrelated gaussian noise, and correlations are accounted for by $\mathcal{C}\mathcal{C}^T$. The matrix $\mathcal{C}$ is thereby not uniquely determined. While $\mathcal{C}\mathcal{C}^T$ is symmetric, $\mathcal{C}$ appears, in general, asymmetric without physical significance; we choose
\begin{equation}
\mathcal{C}= {\scriptsize\frac{\scriptstyle 1}{\scriptstyle 4\sqrt{6N}} \begin{pmatrix} 2\sqrt{\gamma} &  -\sqrt{3}\sqrt{3+2\gamma} &  -\sqrt{3+2\gamma}\\
  2\sqrt{\gamma} & 0 & 2\sqrt{3+2\gamma}\\
 2\sqrt{\gamma} &  \sqrt{3}\sqrt{3+2\gamma} &  -\sqrt{3+2\gamma}   
 \end{pmatrix}}\,.
\label{eq:corr}
\end{equation}

We have numerically solved the SPDE~(\ref{SPDE})  using open software from the XmdS project \cite{xmds}. Snapshots of the resulting steady states, for low diffusivity and  rates $\gamma$ approaching the bifurcation point $\gamma=0$, are shown in Fig.~\ref{snapshots}. All three species coexist in a stable manner, through the formation of spatial patterns. Large values of $\gamma$ (see $\gamma=5$ in Fig.~\ref{snapshots} (a)) yield a dominant contribution of the selection and reproduction events~(\ref{react2}),(\ref{react3}), and entangled rotating spirals form, similar to the structures reported in Refs.~\cite{reichenbach-2007-448,reichenbach-2007-99}. At lower values, these spirals appear blurred, while their wavelength  increases. The blurring intensifies upon lowering $\gamma$; at the bifurcation point ($\gamma=0$) patterns take shape only extremely weakly and appear to be of predominantly stochastic nature.

These observations can be analytically understood by employing a complex Ginzburg-Landau equation (CGLE). Although the RE~(\ref{RE}) operate in a three-dimensional phase space, spanned by the densities $a$, $b$, and $c$, trajectories quickly relax to a two-dimensional invariant manifold~\cite{reichenbach-2007-99}. On the latter,  expanding the RE to third order around the unstable reactive fixed point 
(and ignoring higher nonlinearities) results in the normal form of the Hopf bifurcation~\cite{reichenbach-2007-99}.  The corresponding SPDE, upon ignoring noise, may be cast into the form of a CGLE, where a complex variable $z$ encodes the densities' deviations from the internal fixed point, see Refs.~\cite{reichenbach-2007-99}:
\begin{equation}
\partial_t z=D\Delta z+(c_1-i\omega)z -c_2(1-ic_3)|z|^2z\,.
\label{CGLE}
\end{equation}
The coefficients $\omega$, $c_1$, $c_2$, and $c_3$ are rational functions of the parameter $\gamma$ and given by 
$\omega= \frac{\sqrt{3}}{8}(\gamma +2)$,
$c_1= \frac{\gamma}{8}$, 
$c_2= \gamma\frac{73(7\gamma^2+27\gamma+27)+20\gamma^2}{90(7\gamma^2+27\gamma+27)}$, and
$c_3= \frac{1}{\gamma}\frac{9\sqrt{3}(\gamma+2)\left[23\gamma^2+63(\gamma+1)\right]}{73(7\gamma^2+27\gamma+27)+20\gamma^2}$.
As main characteristics, near the bifurcation point, i.e.~$\gamma\ll 1$, they asymptotically behave as $\omega\approx\text{const.}$, $c_1\sim c_2\sim\gamma$, and $c_3\sim 1/\gamma$.
The vanishing of $c_1$ and $c_2$ at the bifurcation point reflects the neutral stability of the fixed point and the surrounding closed orbits.

\begin{figure}   
\begin{center}    
\includegraphics[scale=1]{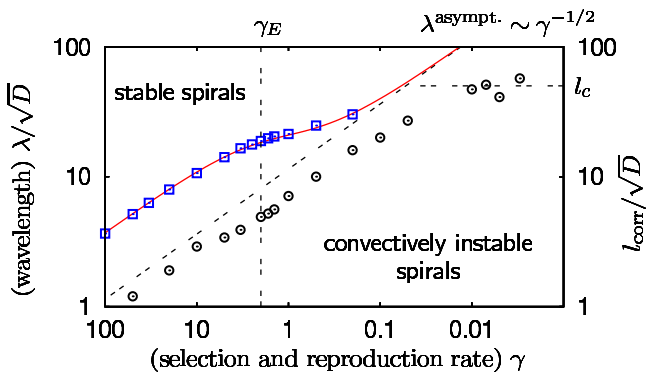}
\vspace*{-0.1cm}
\caption{The dependence of the wavelength $\lambda$ on the rate $\gamma$. Analytical predictions from the CGLE~(\ref{CGLE}) (divided by a factor $1.55$), red line,  are compared to numerical results (\textcolor{blue}{\scriptsize  $\Box$}).  For $\gamma<\gamma_E\approx 2$ spirals are convectively instable. When approaching the bifurcation point, $\gamma\to 0$, computation of  the correlation length ({\scriptsize $\odot$}) shows that the spatial structures are no longer determined by the (diverging) wavelength, but reach a constant size $l_c$, see text. 
\label{gamma_lambda}}
\end{center}                
\end{figure} 
A CGLE is accurate in the vicinity of a supercritical Hopf-bifurcation. In our case, Eq.~(\ref{CGLE}) is only approximate: the RE~(\ref{RE}) do not exhibit a Hopf bifurcation, but a degenerate bifurcation where heteroclinic orbits turn into a family of nestled, neutrally stable orbits. In the following, we show that the CGLE~(\ref{CGLE}) nevertheless provides a reliable description of the system. However, ignoring higher nonlinearities induces certain quantitative deviations from numerical findings, as discussed below.

Rotating spiral waves constitute a generic solution to the CGLE, and the spreading velocities, frequencies, and wavelengths can be calculated analytically~\cite{aranson-2002-74}. As an example, the wavelength follows as
$\lambda=2\pi c_3
\sqrt{D/c_1}
\left(1-\sqrt{1+ c_3^2}\right)^{-1}$.
Comparing these analytic values to numerical ones, we have found that the former exceed the latter by a factor of $1.55$, which we attribute to higher (ignored) nonlinearities. Reducing the analytical result by this constant factor yields an excellent agreement, see Fig.~\ref{gamma_lambda}.

The CGLE~(\ref{CGLE}) predicts an Eckhaus instability. Namely, the spirals are only stable against 
longitudinal long-wave perturbations if the Eckhaus criterion $1+2\frac{1+c_3^2Q^2}{1-Q^2}>0$ is fullfilled, with the rescaled wavevector $Q=2\pi\sqrt{D/c_1}/\lambda$, see Ref.~\cite{aranson-2002-74}. In our case, this condition translates into $\gamma>\gamma_E$, with the value $\gamma_E^\text{anal.}\approx 1.43$. Above $\gamma_E$, the spirals are  absolutely stable, while for values of $\gamma$ below $\gamma_E$, they exhibit convective instabilities: 
a localized perturbation grows but travels away. The instabilities result in the blurring seen in Fig.~\ref{snapshots} and originate in the only weak instability of the RE's internal fixed point. Numerically, the value of $\gamma_E$ may be determined by analyzing the influence of perturbations, as we describe  in the Supplementary Material~\cite{suppl}. A value $\gamma_E\approx2$ is found, which exceeds the analytical one at a factor $1.4$. As for the wavelength, we attribute this deviation to the fact that the CGLE~(\ref{CGLE}) is only a (third-order) approximation to the full nonlinear terms appearing in the SPDE~(\ref{SPDE}).

Upon approaching the bifurcation point, i.e.~when $\gamma\to 0$, the spirals' wavelength, as predicted by the CGLE~(\ref{CGLE}), diverges: To leading order in $1/\gamma$, we calculate $\lambda^\text{asympt.} = 4\pi\sqrt{2D/\gamma}$, such that $\lambda^\text{asympt.}\to\infty$ when $\gamma\to 0$. However, in this limit, the blurring due to the convective instability dominates over the instability of the RE (which indeed vanishes at the bifurcation point), such that the spiral waves are no longer relevant.
Computation of spatial correlation functions~\footnote{e.g. equal-time correlation of species $A$: $g_{AA}(r)=\langle a(0,t)a(r,t)\rangle -  \langle a(0,t)\rangle\langle a(r,t)\rangle$}  and the resulting  correlation length $l_\text{corr}$ (where spatial correlations have decayed to $1/e$ of their maximal value) shows that, for $\gamma<0.001$, $l_\text{corr}$ is no longer proportional to the (diverging) wavelength, but appears to approach a constant value $l_c\sqrt{D}$, see Fig.~\ref{gamma_lambda}. The spatial structures emerging at the bifurcation point are extremely weak, c.f. Fig.~\ref{snapshots} (e), and should be caused by fluctuations alone, as the RE do not yield an instability there. Determining the value $l_c$ requires understanding of  these fluctuation-driven patterns, and may be the subject of future studies.

\begin{figure}   
\begin{center}
\includegraphics[scale=1]{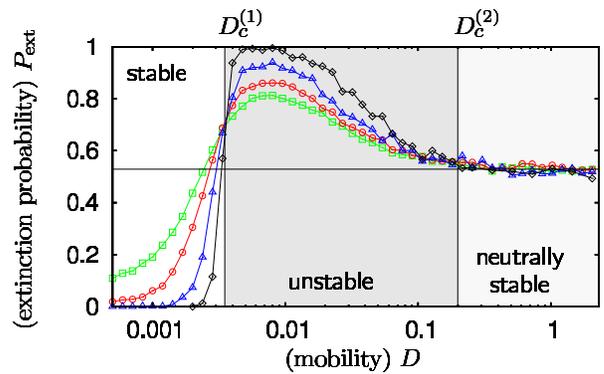} 
\vspace*{-0.1cm}
\caption{Regimes of stable, unstable, and neutrally stable biodiversity. They are revealed by computing the extinction probability 
 for $\gamma=0$ and $\rho=1$, depending on $D$, after a waiting time $t=N$, for increasing system size $N$. We show curves for $N=20\times 20$ (\textcolor{green}{\scriptsize  $\Box$}), $N=25\times 25$ (\textcolor{red}{\large $\circ$}), $N=40\times 40$ (\textcolor{blue}{\scriptsize  $\bigtriangleup$}), and $N=100\times 100$ ({\scriptsize $\diamondsuit$}). The transitions occur at  $D_c^{(1)}\approx 3.5\times 10^{-3}$ and  $D_c^{(2)}\approx 0.2$.
\label{Pext}}
\end{center}                
\end{figure}
For high mobilities, the system becomes effectively well-mixed, and solely the stability  of the RE's internal fixed point determines whether species diversity will be maintained or not~\cite{claussen-2008-100}.  In contrast, when  mobility lies below a threshold value, spatial patterns can form and help to enable stable species diversity~\cite{reichenbach-2007-448}. In the following, we show that, at the bifurcation, patterns can have even different, highly non-trivial impact on  biodiversity.

Fluctuations unavoidably lead to ultimate extinction of species~\cite{reichenbach-2007-448}. However,  transient coexistence can last very long. For this reason, we have proposed a scheme to differentiate stable from unstable diversity which is based on time-scales. In brief, neutral stability leads to a mean extinction time $T$ which is proportional to the system size $N$~\cite{reichenbach-2006-74,antal-2006-68}. We therefore consider the asymptotic limit $N\to\infty$, and define a situation where $T/N\to\infty$ as stable diversity (extinction takes very long), while $T/N\to 0$ coresponds to unstable diversity (extinction is fast).

We have applied this concept to determine the influence of mobility on diversity's stability at the bifurcation point $\gamma=0$. From stochastic simulations of the reactions~(\ref{react1}) with nearest-neighbor exchanges on lattices of increasing size $N$, we have computed the extinction probability $P_\text{ext}$ that, starting at a random initial state with equal densities of $A$, $B$, and $C$, two species have gone extinct after a waiting time $t=N$.  Results are shown in Fig.~\ref{Pext}, where three different mobility regimes emerge. For low mobilities (around $D=0.001$), $P_\text{ext}$ approaches $0$ for increasing $N$, implying $T/N\to\infty$. In this regime, coexistence is therefore stable, 
as has previously been observed for vanishing mobility, see~\cite{szabo-2007-446} and Refs. therein. For intermediate mobilities around $D=0.007$, the opposite behavior emerges:  $P_\text{ext}$ tends to $1$, such that $T/N\to 0$ and coexistence is unstable. High mobilities, around $D=0.5$, yield an asymptotic value of $P_\text{ext}$ of about $0.53$,
independent of $N$,  and thus neutral stability~\cite{reichenbach-2007-448} (the precise value of $P_\text{ext}$  depends on the choice of $t\sim N$).
Critical mobility values separate these three regimes. Namely, with increasing $N$, the transition from the stable to the unstable regime becomes sharp at a value $D_c^{(1)}\approx 3.5\times 10^{-3}$. The  unstable regime verges on the neutrally stable one around $D_c^{(2)}\approx 0.2$, 
where $P_\text{ext}$ reaches $0.53$. Our data do not reveal whether a sharp transition or a crossover results when $N\to\infty$.

In summary, we have quantitatively analyzed the emergence of blurred (convectively instable) spiral waves near a bifurcation point of cyclic dynamics. This bifurcation point is characterized by neutrally stable, cycling orbits predicted by the corresponding RE~(\ref{RE}).  There, spatial structures are predominantly determined by noise.  These patterns have ambiguous impact on maintaining the coexistence of the interacting species. Low individuals' mobility, corresponding to small patterns, promotes diversity, as has already been observed in previous studies~\cite{kerr-2002-418,szabo-2007-446, reichenbach-2007-448}. However, medium values of mobility, inducing relatively large patterns, lead to rapid species extinction. Only for high mobilities, spatial patterns have no influence, and, being at the bifurcation point, neutrally stable coexistence emerges. Further investigations of the destabilizing influence of spatial patterns at medium mobilities are required for a general understanding of the effects of spatial degrees of freedom on the coexistence of mobile individuals. Also, such studies will shed further light on the role of noise that becomes a major player in pattern formation at  bifurcation points.

Financial support of the German Excellence Initiative via the program ``Nanosystems
Initiative Munich" and the German Research Foundation via the SFB TR12 ``Symmetries and Universalities in Mesoscopic Systems''  is gratefully acknowledged. T. R. acknowledges funding by the Elite-Netzwerk Bayern.



\begin{thebibliography}{27}
\expandafter\ifx\csname natexlab\endcsname\relax\def\natexlab#1{#1}\fi
\expandafter\ifx\csname bibnamefont\endcsname\relax
  \def\bibnamefont#1{#1}\fi
\expandafter\ifx\csname bibfnamefont\endcsname\relax
  \def\bibfnamefont#1{#1}\fi
\expandafter\ifx\csname citenamefont\endcsname\relax
  \def\citenamefont#1{#1}\fi
\expandafter\ifx\csname url\endcsname\relax
  \def\url#1{\texttt{#1}}\fi
\expandafter\ifx\csname urlprefix\endcsname\relax\def\urlprefix{URL }\fi
\providecommand{\bibinfo}[2]{#2}
\providecommand{\eprint}[2][]{\url{#2}}

\bibitem[{\citenamefont{May}(1974)}]{May}
\bibinfo{author}{\bibfnamefont{R.~M.} \bibnamefont{May}},
  \emph{\bibinfo{title}{Stability and complexity in model ecosystems}}
  (\bibinfo{publisher}{Princeton Univ. Press}, \bibinfo{year}{1974}),
  \bibinfo{edition}{2nd} ed.

\bibitem[{\citenamefont{Levin}(1974)}]{levin-1974-108}
\bibinfo{author}{\bibfnamefont{S.~A.} \bibnamefont{Levin}},
  \bibinfo{journal}{Am. Nat.} \textbf{\bibinfo{volume}{108}},
 \bibinfo{pages}{207} (\bibinfo{year}{1974}).

\bibitem[{\citenamefont{Hassell et~al.}(1994)\citenamefont{Hassell, Comins, and
  May}}]{hassell-1994-370}
\bibinfo{author}{\bibfnamefont{M.~P.} \bibnamefont{Hassell}},
  \bibinfo{author}{\bibfnamefont{H.~N.} \bibnamefont{Comins}},
  \bibnamefont{and} \bibinfo{author}{\bibfnamefont{R.~M.} \bibnamefont{May}},
  \bibinfo{journal}{Nature} \textbf{\bibinfo{volume}{370}},
  \bibinfo{pages}{290} (\bibinfo{year}{1994}).

\bibitem[{\citenamefont{Durrett and Levin}(1994)}]{durrett-1994-46}
\bibinfo{author}{\bibfnamefont{R.}~\bibnamefont{Durrett}} \bibnamefont{and}
  \bibinfo{author}{\bibfnamefont{S.}~\bibnamefont{Levin}},
  \bibinfo{journal}{Theor. Pop. Biol.} \textbf{\bibinfo{volume}{46}},
  \bibinfo{pages}{363} (\bibinfo{year}{1994}).

\bibitem[{\citenamefont{Reichenbach
  et~al.}(2007{\natexlab{a}})\citenamefont{Reichenbach, Mobilia, and
  Frey}}]{reichenbach-2007-448}
\bibinfo{author}{\bibfnamefont{T.}~\bibnamefont{Reichenbach}},
  \bibinfo{author}{\bibfnamefont{M.}~\bibnamefont{Mobilia}}, \bibnamefont{and}
  \bibinfo{author}{\bibfnamefont{E.}~\bibnamefont{Frey}},
  \bibinfo{journal}{Nature} \textbf{\bibinfo{volume}{448}},
  \bibinfo{pages}{1046} (\bibinfo{year}{2007}{\natexlab{a}}).

\bibitem[{\citenamefont{Kerr et~al.}(2002)\citenamefont{Kerr, Riley, Feldman,
  and Bohannan}}]{kerr-2002-418}
\bibinfo{author}{\bibfnamefont{B.}~\bibnamefont{Kerr et al.}},
 \bibinfo{journal}{Nature}
  \textbf{\bibinfo{volume}{418}}, \bibinfo{pages}{171} (\bibinfo{year}{2002}).

\bibitem[{\citenamefont{Sinervo and Lively}(1996)}]{sinervo-1996-340}
\bibinfo{author}{\bibfnamefont{B.}~\bibnamefont{Sinervo}} \bibnamefont{and}
  \bibinfo{author}{\bibfnamefont{C.~M.} \bibnamefont{Lively}},
  \bibinfo{journal}{Nature} \textbf{\bibinfo{volume}{380}},
  \bibinfo{pages}{240} (\bibinfo{year}{1996}).

\bibitem[{\citenamefont{Jackson and Buss}(1975)}]{jackson-1975-72}
\bibinfo{author}{\bibfnamefont{J.~B.~C.} \bibnamefont{Jackson}}
  \bibnamefont{and} \bibinfo{author}{\bibfnamefont{L.}~\bibnamefont{Buss}},
  \bibinfo{journal}{Proc. Natl. Acad. Sci. U.S.A.}
  \textbf{\bibinfo{volume}{72}}, \bibinfo{pages}{5160} (\bibinfo{year}{1975}).

\bibitem[{\citenamefont{Szab\'o and Szolnoki}(2002)}]{szabo-2002-65}
\bibinfo{author}{\bibfnamefont{G.}~\bibnamefont{Szab\'o}} \bibnamefont{and}
  \bibinfo{author}{\bibfnamefont{A.}~\bibnamefont{Szolnoki}},
  \bibinfo{journal}{Phys. Rev. E} \textbf{\bibinfo{volume}{65}},
  \bibinfo{pages}{036115} (\bibinfo{year}{2002}).

\bibitem[{\citenamefont{Mobilia et~al.}(2006)\citenamefont{Mobilia, Georgiev,
  and T\"auber}}]{mobilia-2006-73}
\bibinfo{author}{\bibfnamefont{M.}~\bibnamefont{Mobilia}},
  \bibinfo{author}{\bibfnamefont{I.~T.} \bibnamefont{Georgiev}},
  \bibnamefont{and} \bibinfo{author}{\bibfnamefont{U.~C.}
  \bibnamefont{T\"auber}}, \bibinfo{journal}{Phys. Rev. E}
  \textbf{\bibinfo{volume}{73}}, \bibinfo{pages}{040903(R)}
  (\bibinfo{year}{2006}).

\bibitem[{\citenamefont{Szab\'o and Fath}(2007)}]{szabo-2007-446}
\bibinfo{author}{\bibfnamefont{G.}~\bibnamefont{Szab\'o}} \bibnamefont{and}
  \bibinfo{author}{\bibfnamefont{G.}~\bibnamefont{Fath}},
  \bibinfo{journal}{Phys. Rep.} \textbf{\bibinfo{volume}{446}},
  \bibinfo{pages}{97} (\bibinfo{year}{2007}).

\bibitem[{\citenamefont{May and Leonard}(1975)}]{may-1975-29}
\bibinfo{author}{\bibfnamefont{R.~M.} \bibnamefont{May}} \bibnamefont{and}
  \bibinfo{author}{\bibfnamefont{W.~J.} \bibnamefont{Leonard}},
  \bibinfo{journal}{SIAM J. Appl. Math} \textbf{\bibinfo{volume}{29}},
  \bibinfo{pages}{243} (\bibinfo{year}{1975}).

\bibitem[{\citenamefont{Busse and Heikes}(1980)}]{busse-1980-208}
\bibinfo{author}{\bibfnamefont{F.~H.} \bibnamefont{Busse}} \bibnamefont{and}
  \bibinfo{author}{\bibfnamefont{K.~E.} \bibnamefont{Heikes}},
  \bibinfo{journal}{Science} \textbf{\bibinfo{volume}{208}},
  \bibinfo{pages}{173} (\bibinfo{year}{1980}).

\bibitem[{\citenamefont{Tu and Cross}(1992)}]{yuhai-1992-69}
\bibinfo{author}{\bibfnamefont{Y.}~\bibnamefont{Tu}} \bibnamefont{and}
  \bibinfo{author}{\bibfnamefont{M.~C.} \bibnamefont{Cross}},
  \bibinfo{journal}{Phys. Rev. Lett.} \textbf{\bibinfo{volume}{69}},
  \bibinfo{pages}{2515} (\bibinfo{year}{1992}).

\bibitem[{\citenamefont{Reichenbach et~al.}(2006)\citenamefont{Reichenbach,
  Mobilia, and Frey}}]{reichenbach-2006-74}
\bibinfo{author}{\bibfnamefont{T.}~\bibnamefont{Reichenbach}},
  \bibinfo{author}{\bibfnamefont{M.}~\bibnamefont{Mobilia}}, \bibnamefont{and}
  \bibinfo{author}{\bibfnamefont{E.}~\bibnamefont{Frey}},
  \bibinfo{journal}{Phys. Rev. E} \textbf{\bibinfo{volume}{74}},
  \bibinfo{pages}{051907} (\bibinfo{year}{2006}).

\bibitem[{\citenamefont{Redner}(1983)}]{Redner}
\bibinfo{author}{\bibfnamefont{S.}~\bibnamefont{Redner}},
  \emph{\bibinfo{title}{A guide to first-passage processes}}
  (\bibinfo{publisher}{Cambridge University Press}, \bibinfo{year}{1983}).



\bibitem[{\citenamefont{Reichenbach
  et~al.}(2007{\natexlab{b}})\citenamefont{Reichenbach, Mobilia, and
  Frey}}]{reichenbach-2007-99}
\bibinfo{author}{\bibfnamefont{T.}~\bibnamefont{Reichenbach}},
  \bibinfo{author}{\bibfnamefont{M.}~\bibnamefont{Mobilia}}, \bibnamefont{and}
  \bibinfo{author}{\bibfnamefont{E.}~\bibnamefont{Frey}},
  \bibinfo{journal}{Phys. Rev. Lett.} \textbf{\bibinfo{volume}{99}},
  \bibinfo{pages}{238105} (\bibinfo{year}{2007}{\natexlab{b}});
\bibinfo{author}{\bibfnamefont{T.}~\bibnamefont{Reichenbach}},
  \bibinfo{author}{\bibfnamefont{M.}~\bibnamefont{Mobilia}}, \bibnamefont{and}
  \bibinfo{author}{\bibfnamefont{E.}~\bibnamefont{Frey}},
  (\bibinfo{year}{2008}), \bibinfo{note}{JTB in print}.



\bibitem[{\citenamefont{Gardiner}(1983)}]{Gardiner}
\bibinfo{author}{\bibfnamefont{C.~W.} \bibnamefont{Gardiner}},
  \emph{\bibinfo{title}{Handbook of Stochastic Methods}}
  (\bibinfo{publisher}{Springer}, \bibinfo{year}{1983}), \bibinfo{edition}{1st}
  ed.



\bibitem[{\citenamefont{Traulsen et~al.}(2005)\citenamefont{Traulsen, Claussen,
  and Hauert}}]{traulsen-2005-95}
\bibinfo{author}{\bibfnamefont{A.}~\bibnamefont{Traulsen}},
  \bibinfo{author}{\bibfnamefont{J.~C.} \bibnamefont{Claussen}},
  \bibnamefont{and} \bibinfo{author}{\bibfnamefont{C.}~\bibnamefont{Hauert}},
  \bibinfo{journal}{Phys. Rev. Lett.} \textbf{\bibinfo{volume}{95}},
  \bibinfo{pages}{238701} (\bibinfo{year}{2005}).

\bibitem[{XmdS()}]{xmds}
XmdS, \bibinfo{note}{\texttt{http://www.xmds.org}}.

\bibitem[{\citenamefont{Aranson and Kramer}(2002)}]{aranson-2002-74}
\bibinfo{author}{\bibfnamefont{I.~S.} \bibnamefont{Aranson}} \bibnamefont{and}
  \bibinfo{author}{\bibfnamefont{L.}~\bibnamefont{Kramer}},
  \bibinfo{journal}{Rev. Mod. Phys.} \textbf{\bibinfo{volume}{74}},
  \bibinfo{pages}{99} (\bibinfo{year}{2002});
\bibinfo{author}{\bibfnamefont{W.}~\bibnamefont{van Saarloos}},
  \bibinfo{journal}{Phys. Rep.} \textbf{\bibinfo{volume}{386}},
  \bibinfo{pages}{29} (\bibinfo{year}{2003}).

\bibitem[{sup()}]{suppl}
\bibinfo{note}{Supplementary EPAPS Document}.

\bibitem[{\citenamefont{Claussen and Traulsen}(2008)}]{claussen-2008-100}
\bibinfo{author}{\bibfnamefont{J.~C.} \bibnamefont{Claussen}} \bibnamefont{and}
  \bibinfo{author}{\bibfnamefont{A.}~\bibnamefont{Traulsen}},
  \bibinfo{journal}{Phys. Rev. Lett.} \textbf{\bibinfo{volume}{100}},
  \bibinfo{pages}{058104} (\bibinfo{year}{2008}).

\bibitem[{\citenamefont{Antal and Scheuring}(2006)}]{antal-2006-68}
\bibinfo{author}{\bibfnamefont{T.}~\bibnamefont{Antal}} \bibnamefont{and}
  \bibinfo{author}{\bibfnamefont{I.}~\bibnamefont{Scheuring}},
  \bibinfo{journal}{Bull. Math. Biol.} \textbf{\bibinfo{volume}{68}},
  \bibinfo{pages}{1923} (\bibinfo{year}{2006}).

\end{thebibliography}

\newpage

\onecolumngrid

\setcounter{page}{1}

\begin{center}
{ \bf \Large Supplementary Material}\\
\vspace*{0.5cm}
\large Numerical determination of the onset of convective instability
\end{center}
\vspace*{0.5cm}

In this Supplementary Material, we describe further how the convective instability takes shape in the  system's reactive steady state, and how one can therefrom infer numerically the value $\gamma_E$, i.e.~the  boundary between stable and convectively unstable spiral waves.

We have already seen that convective instability manifests in a blurring of the spirals. However, at the onset of convective instability, the blurring is extremely weak, and cannot be detected from analyzing snapshots. Moreover, further inside the region of convective instability, where the blurring is stronger, it can hardly be quantified using, e.g., correlation functions. For these reasons, the blurring seems not to be a good candidate for a quantification of the onset of convective instability.

A better suited measure is obtained by investigating the effects of convective instability on large spiral waves. To obtain the latter, we solve the  SPDE~(\ref{SPDE}) with very low noise, but starting from an initial state which is spatially inhomogeneous. The latter inhomogeneities serve as ``seeds'' for the  developing spiral waves and determine the position and numbers of their vortices~\cite{reichenbach-2007-99}. In our numerical solutions, we have started from initial densities $a({\bf r},t=0)= a^*+1/100 \times\cos(2\pi r_1r_2),~b({\bf r},t=0)=b^*,~c({\bf r},t=0)=c^*$. Hereby, $a^*,b^*,c^*$ denote the densities at the reactive fixed point of the RE~(\ref{RE}).

\begin{figure*}[h]  
\begin{center}
\begin{tabular}{lllll}
{ (a)} \hspace{0.8cm} $\gamma=100$ &  { (b)} \hspace{0.8cm} $\gamma=5$ & {(c)} \hspace{0.8cm} $\gamma=2.5$  & {(d)} \hspace{0.8cm} $\gamma=2$  \\
\includegraphics[scale=0.38]{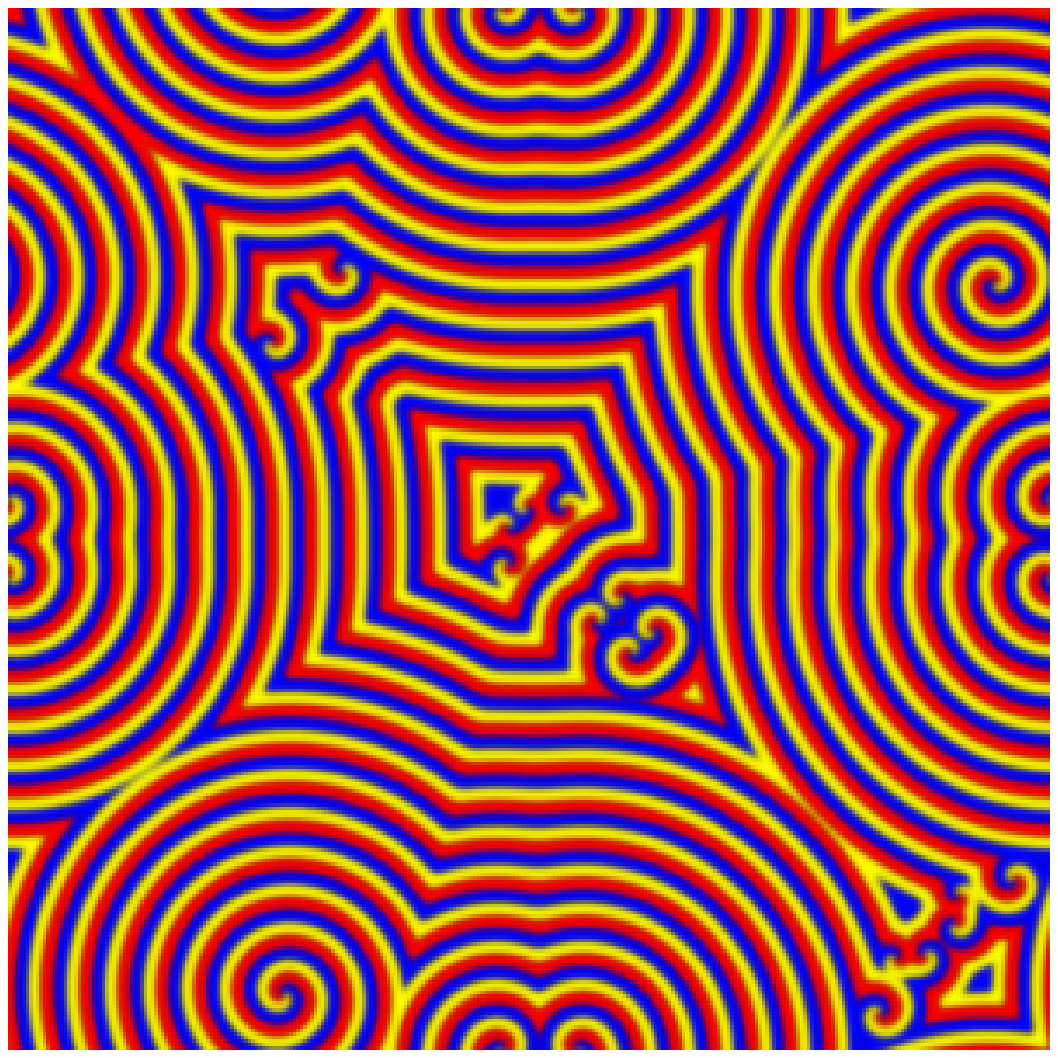}&
\includegraphics[scale=0.38]{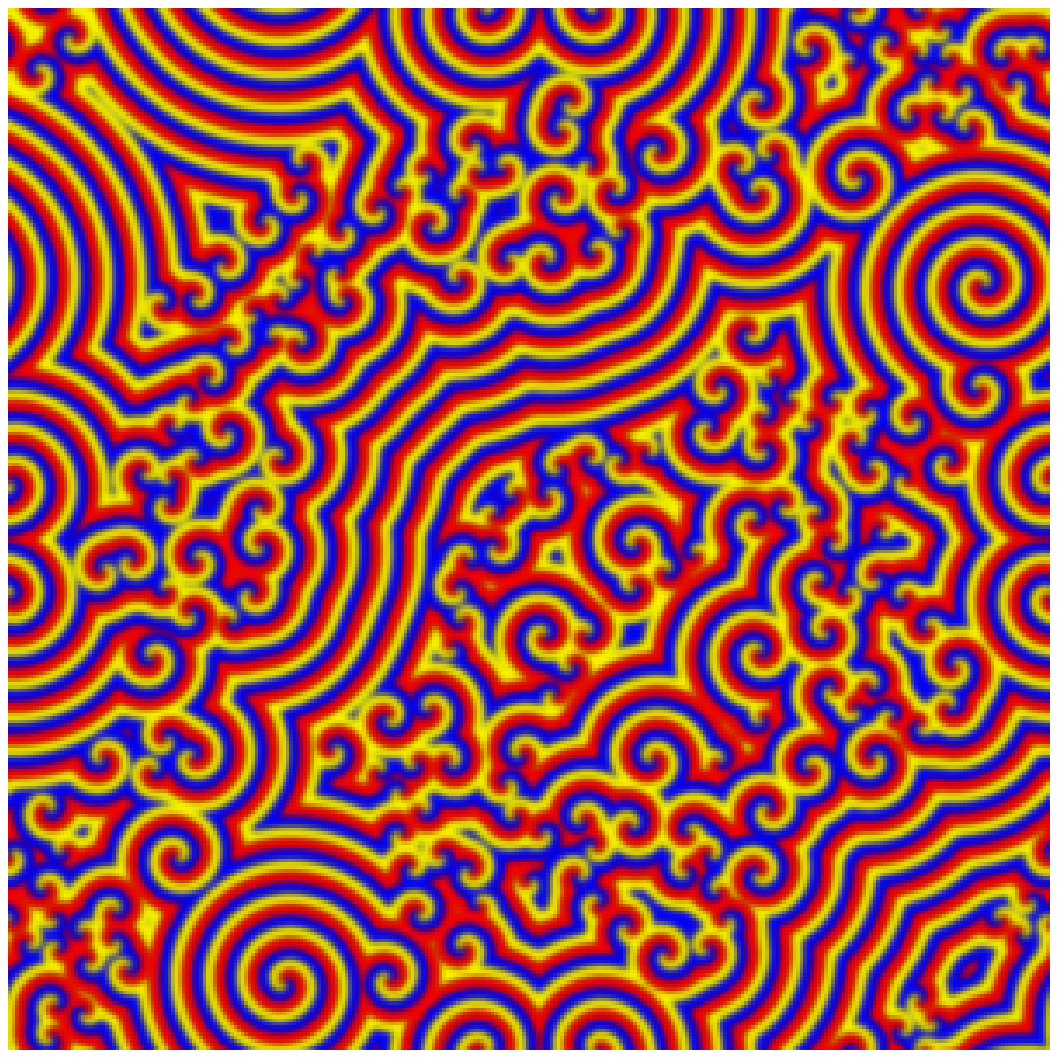}&
\includegraphics[scale=0.38]{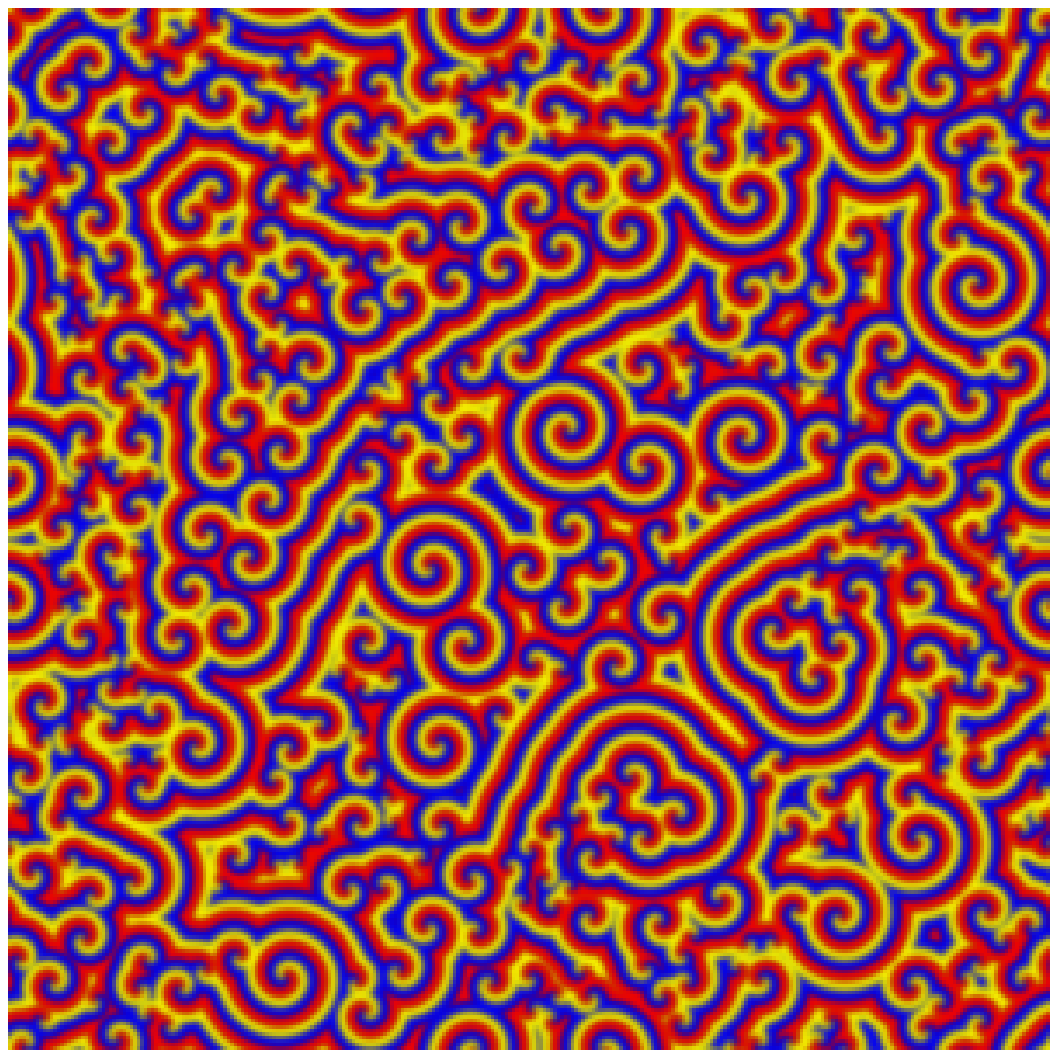}&
\includegraphics[scale=0.38]{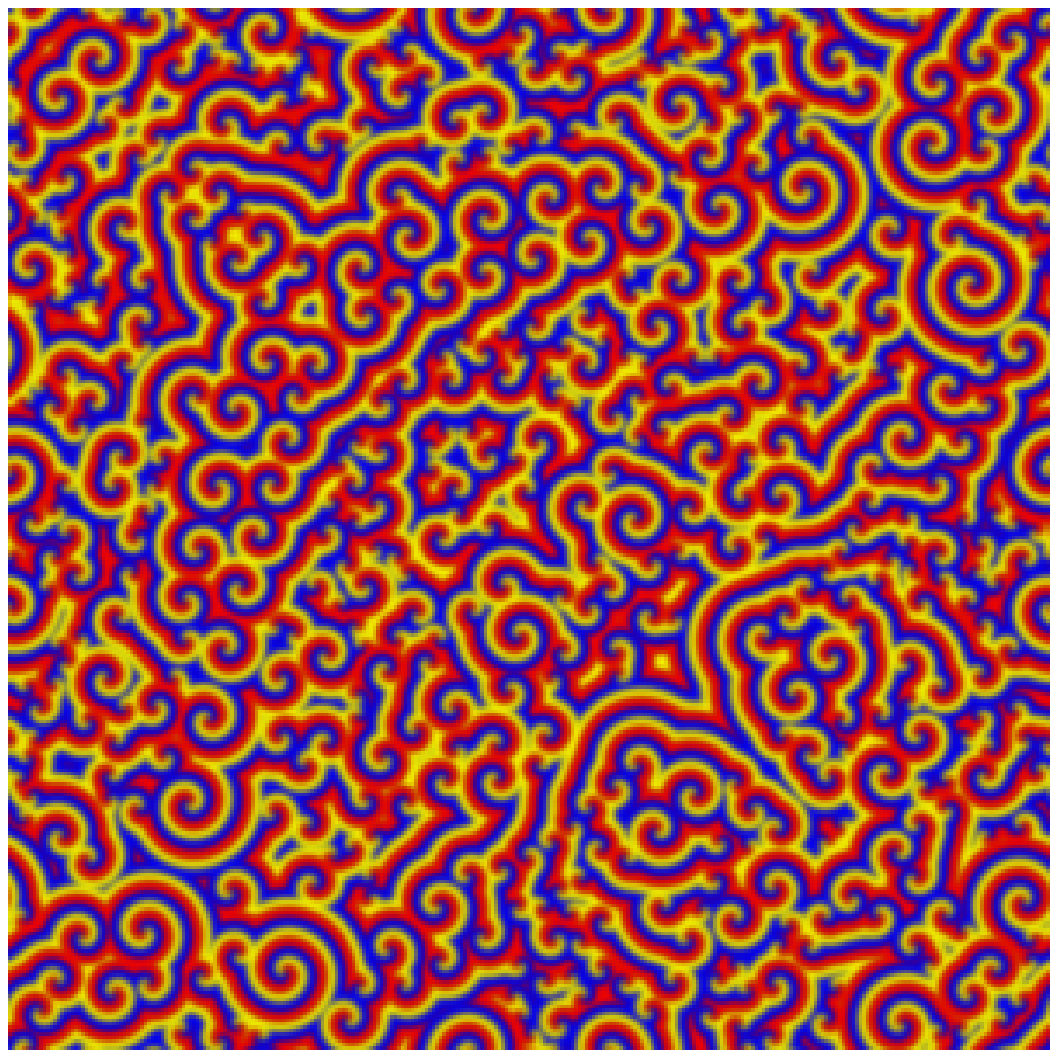}
\end{tabular}
\caption{Snapshots of the reactive steady state for different values of the rate $\gamma$, and a diffusivity of $D=10^{-6}\times\gamma$. We have solved the deterministic PDE, meaning Eqs.~(\ref{SPDE}) in the absence of noise, with initial spatial inhomogeneities: $a({\bf r},t=0)=a^*+1/100\cos(2\pi r_1r_2),~b({\bf r},t=0)=b^*,~c({\bf r},t=0)=c^*$. From left to right, the value of $\gamma$ decreases, approaching the Eckhaus instability. At the same time, the spiral density rises.\label{snapshots_inst}}
\end{center}                
\end{figure*} 
Deep in the regime of stable spiral waves, i.e.~for $\gamma \gg \gamma_E$, few large spirals emerge from these initial inhomogeneities, see Fig.~\ref{snapshots_inst} (a). For decreasing values of $\gamma$, approaching the convective instability, more and more vortices appear, see Fig.~\ref{snapshots_inst} (b),(c). Apparently, upon approaching the onset of convective instability, the large spiral waves become increasingly unstable against perturbations and noise, such that a larger number of smaller spirals appears. The density of spiral vortices increases, upon saturating at a constant  value of about $0.3$ for $\gamma \leq 2$. Interestingly, the same spiral density also emerges when the SPDE~(\ref{SPDE}) are solved starting without initial inhomogeneities, such that noise perturbs the system initially and leads to the emergence of entangled spiral waves. Consequently, at the onset of the convective instability (as well as in the regime of convective instability), upon starting from initial spatial inhomogeneities or employing noise as an internal source of inhomogeneities, the system reaches statistically equivalent states. This is not true within the regime of stable spirals: There, initially imposed spatial inhomogeneities can lead to very different vortex densities than emerge when fluctuations determine the formation of entangled spirals.

Different vortex densities manifest in differences in the spatial correlation functions. The equal-time autocorrelation $g_{aa}(r)$ of species $A$ at distance $r$  shows damped oscillations, and may be approximated by
\begin{equation}
g_{aa}(r)\approx a\exp(-r/b)J_0(r/c)\,,
\end{equation}
with constants $a,b,c$. Hereby, the Bessel function $J_0$ describes oscillations, at a length proportional to $b$, which relate to the wavelength of the spirals. The exponential damping  $\exp(-r/b)$ stems from the entangled spiral waves. The typical length $b$, at which the oscillations decay, reveals information about the vortex density: $b$ is proportional to the mean distance between spiral vortices, and therefore proportional to the inverse square root of the vortex density.

\begin{figure}[t]   
\begin{center}    
\psfrag{lambda}{\begin{rotate}{90}\hspace{-1.9cm}(corr. decay length) $b/\sqrt{D}$\end{rotate}}
\psfrag{sigmac}{\hspace{-0.1cm}$\gamma_E$}
\psfrag{sigma}{\hspace{-1cm}(selection and reproduction rate) $\gamma$}
\psfrag{lambdaneutral}{}
\psfrag{stable}{\hspace{-0.25cm}\sf stable spirals} 
\psfrag{convective}{\hspace{-0.5cm}\parbox{1.5cm}{\vspace{0.32cm}\sf convectively instable spirals}}
\psfrag{corr}{\begin{rotate}{90}\hspace{-1cm}  $l_\text{corr}/\sqrt{D}$\end{rotate}}
\psfrag{lf}{$l_c$}
\psfrag{gamma12}{$\gamma^{-1/2}$}
\includegraphics[scale=0.5]{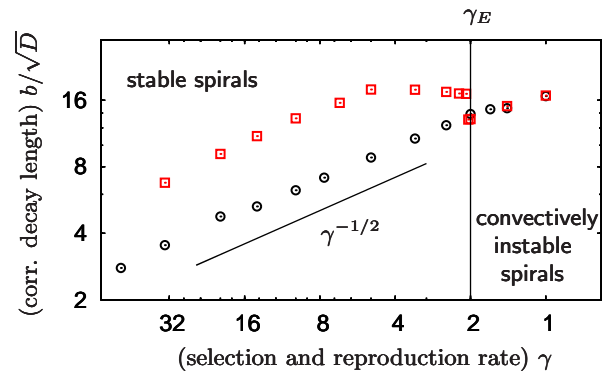}
\vspace*{0.3cm}
\caption{The decay length $b$ of the oscillations of the correlation functions versus the rate $\gamma$. We show results from solving the SPDE~(\ref{SPDE}) both starting from spatial inhomogeneities (red squares) as well as from initially homogeneous states (black circles). In the regime of stable spirals, the decay length is larger for the  solution from initial inhomogeneities, corresponding to larger spirals. Approaching the Eckhaus instability, the decay length approaches the values of the initially homogeneous system. As a remark, similar to the wavelength, $b$ is asymptotically (for $\gamma\gg\gamma_E$ as well as for $\gamma\ll\gamma_E$) proportional to $\gamma^{-1/2}$. The data have been obtained by numerical solution of the SPDE~(\ref{SPDE}) with diffusivity $D=10^{-5}\times\gamma$.}
\end{center}                
\end{figure}

From numerical simulations, obtained from initial inhomogeneities, we have determined the value of $b$ for different rates $\gamma$. As expected, for large values of $\gamma$,  $\gamma\gg\gamma_E$, i.e.~in the regime of stable spirals, $b$ is large, corresponding to a small vortex density arising from the small density of initially imposed spatial inhomogeneities. Decreasing $\gamma$, the value of $b$ decreases, and for $\gamma\leq\gamma_E\approx 2$ reaches  the same value as arises when stochastic effects induce spiral waves.

\end{document}